\documentclass[aps,prc,letterpaper,11pt,twoside,tightenlines,nofootinbib,showpacs,preprint]{revtex4-1}
\usepackage{graphicx}
\usepackage{epsfig,float}
\usepackage[sort&compress]{natbib}
\usepackage{subfigure}
\usepackage{amsmath}
\usepackage{amsfonts}
\usepackage{cancel}
\usepackage{amssymb}
\usepackage{hyperref}
\usepackage{multirow}

\usepackage{color}

\begin{document}
\graphicspath{{./figure/}}

\title{Description of $\rho (1700)$ as a $\rho K \bar{K}$ system with the fixed center approximation}

\author{M.~Bayar$^{1,2}$, W.~H.~Liang$^{1,3}$, T.~Uchino$^{1}$, C.~W.~Xiao$^{1}$}
\affiliation{
$^1$Departamento de F{\'i}sica Te{\'o}rica and IFIC,
Centro Mixto Universidad de Valencia-CSIC,
Institutos de Investigaci{\'o}n de Paterna,
Apartado 22085, 46071 Valencia, Spain\\
\\
$^2$ Department of Physics, Kocaeli University, 41380, Izmit, Turkey\\
\\
$^3$ Department of Physics, Guangxi Normal University, Guilin, 541004, P. R. China
}

\date{\today}

\begin{abstract}
We study the $\rho K\bar{K}$ system with an aim to describe the $\rho (1700)$ resonance.
The chiral unitary approach has achieved success in a description of systems of the light hadron sector.
With this method, the $K \bar{K}$ system in the isospin sector $I=0$, is found to be a dominant component of the $f_0 (980)$ resonance.
Therefore, by regarding the $K\bar{K}$ system as a cluster, the $f_0 (980)$ resonance, we evaluate the $\rho K\bar{K}$ system applying the fixed center approximation to the Faddeev equations.
We construct the $\rho K$ unitarized amplitude using the chiral unitary approach.
As a result, we find a peak in the three-body amplitude around 1739 MeV and a width of about 227 MeV.
The effect of the width of $\rho$ and $f_0 (980)$ is also discussed.
We associate this peak to the $\rho (1700)$ which has a mass of $1720 \pm 20$ MeV and a width of $250 \pm 100$ MeV.
\end{abstract}

\maketitle

\section{Introduction}
One of the most important aims of hadron physics is getting better understanding of the strong interaction through the study of hadronic resonances.
Besides the constituent quark model, recently the rich spectrum of hadronic resonances is studied actively from various viewpoints.
Among them, hadronic molecules, dynamically generated states through the hadronic interaction, attracts plenty of attention.
At low energies, the dynamics of light hadrons can be described in terms of chiral symmetry \cite{Gasser:1984gg,Meissner:1993ah,Pich:1995bw,Ecker:1994gg,Bernard:1995dp}.
By using the leading order of the chiral Lagrangians as input, a powerful tool, the chiral unitary approach, implementing unitarity in coupled channels, has been developed and it has provided great success in describing many resonances for meson-meson or meson-baryon systems \cite{Kaiser:1995eg,Oller:1997ti,Oset:1997it,Oller:1998hw,Oller:1998zr,Oller:2000fj,Jido:2003cb,Guo:2006fu,GarciaRecio:2002td,GarciaRecio:2005hy,Hyodo:2002pk} .

In order to explore multi-hadron systems, the application of the fixed center approximation to Faddeev equations has been implemented \cite{Faddeev:1960su,Toker:1981zh,Barrett:1999cw,Deloff:1999gc,Kamalov:2000iy,Gal:2006cw}.
Under the condition where the cluster structure in three-body systems is not varied so much against the collision of the other particle, that approximation will work fine, as discussed in refs.~\cite{MartinezTorres:2010ax,Bayar:2011qj}.
Moreover as studied in refs.~\cite{Roca:2010tf,YamagataSekihara:2010qk,Xiao:2012dw}, one can proceed to multi-hadron systems with more than three hadrons.

In the present paper, we study the $\rho K\bar{K}$ system in the sector $I(J^P)=1(1^-)$ within the fixed center approximation to obtain the $\rho (1700)$ resonance.
Namely, a pair of $K \bar{K}$ is assumed to form the scalar cluster, the $f_0 (980)$ resonance.
Since the $K \bar{K}$ component in $f_0(980)$ is found to be dominant \cite{Oller:1997ti}, this assumption seems to work well.
The amplitude needed in the present work is the $\rho K$ unitarized scattering amplitude.
In ref.~\cite{Birse:1996hd}, interactions including vector mesons are reviewed.
In order to obtain the $\rho K$ amplitude, we follow the schemes given by refs.~\cite{Roca:2005nm,Geng:2006yb} and extend them to the isospin $I=3/2$ sector.

The $\rho K \bar{K}$ three-body amplitude is constructed under the fixed center approximation.
Basically we follow the formalism given by refs.~\cite{Roca:2010tf,YamagataSekihara:2010qk,Xiao:2012dw,Liang:2013yta}.
In this paper, due to the symmetry of the $\rho K$ and $\rho \bar{K}$ interactions, the formalism is further simplified with respect to ref.~\cite{Liang:2013yta}.

\section{\label{sec:rhoK} $\rho K$ unitarized amplitude}

To implement the Faddeev equation within the fixed center approximation, we need the two-body unitarized amplitude.
Namely in the present case of the $\rho K \bar{K}$ system, the $\rho K~(\rho \bar{K})$ unitarized amplitude is necessary.
In the previous work \cite{Roca:2005nm,Geng:2006yb}, the vector-pseudoscalar interaction in the sector with strangeness $S=1$ and isospin $I=1/2$ was studied within the framework of the chiral unitary approach and that interaction was shown to generate two resonance poles corresponding to the $K_1 (1270)$ resonance.
Here we are going to follow the scheme of refs.~\cite{Roca:2005nm,Geng:2006yb} and pick up only the essence for simplicity.

Following the Bethe-Salpeter approach, we have the $VP$ two-body scattering amplitude as
  \begin{eqnarray}
  T
  =
  [1+V\hat{G}]^{-1} (-V) \vec{\epsilon} \cdot \vec{\epsilon}',
  \label{eq:rhoK_amp}
  \end{eqnarray}
where $V$ is an interaction kernel which will be discussed later, $\hat{G}$ is $(1+ \frac{1}{3}\frac{q_l^2}{M_l^2})G$ being a diagonal matrix and $\vec{\epsilon} (\vec{\epsilon}') $ represents a polarization vector of the incoming (outgoing) vector-meson.
Thanks to the on-shell factorization, a loop function of pseudoscalar and vector mesons $G_l$ can be expressed as a function of the energy $\sqrt{s}$
  \begin{eqnarray}
  G_l(\sqrt{s})
  =
  i \int \frac{d^4q}{(2\pi)^4}
  \frac{1}{(P-q)^2 - M_l^2 + i\epsilon}
  \frac{1}{q^2 - m_l^2 + i\epsilon} , \nonumber \\
  \end{eqnarray}
where $m_l$ and $M_l$ are masses of pseudoscalar and vector in the $l$th channel respectively and the four dimensional momentum $P$ is determined at the rest frame,  $P=(\sqrt{s},\vec{0})$.
To remove the ultra violet divergence of the loop function, we follow the dimensional regularization scheme
  \begin{eqnarray}
  G_l(\sqrt{s})
  &=&
  \frac{1}{16\pi^2}
  \left\{
  a(\mu) + {\rm ln} \frac{M_l^2}{\mu^2} + \frac{m_l^2 - M_l^2 + s}{2s} {\rm ln}\frac{m_l^2}{M_l^2}
  \right. \nonumber \\
  &&
  + \frac{q_l}{\sqrt{s}}
  \left[
  {\rm ln} (s - (M_l^2 - m_l^2) + 2q_l \sqrt{s})
  \right. \nonumber \\
  &&
  + {\rm ln} (s + (M_l^2 - m_l^2) + 2q_l \sqrt{s})
  \nonumber \\
  &&
  -{\rm ln} (-s + (M_l^2 - m_l^2) + 2q_l \sqrt{s})
  \nonumber \\
  &&
  \left.
  \left.
  -{\rm ln} (-s - (M_l^2 - m_l^2) + 2q_l \sqrt{s})
  \right]
  \right\}
  \label{eq:G_rhoK},
  \end{eqnarray}
with a momentum $q_l$ determined at the center of mass frame
  \begin{eqnarray}
  q_l
  =
  \frac{ \sqrt{ [s- (M-m)^2][s - (M+m)^2)] }}{ 2\sqrt{s} },
  \end{eqnarray}
where $\mu$ is a scale parameter in this scheme.
The finite part of the loop function is stable against changes of $\mu$ due to the subtraction constant $a(\mu)$ which absorbs the changes of $\mu$.

Before the derivation of the $VP$ interaction, it is worth referring to a finite width of the vector mesons in the loop function.
In ref.~\cite{Geng:2006yb}, the effect of the propagation of unstable particles is taken into account in terms of the Lehmann representation.
That is done with the dispersion relation with its imaginary part
  \begin{eqnarray}
  D(s)
  =
  \int_{ s_{ {\rm th} }}^{\infty}
  ds_V
  \left( - \frac{1}{\pi} \right)
  \frac{{\rm Im} D(s_V)}{s - s_V + i\epsilon }
  \label{eq:disp},
  \end{eqnarray}
where $s_{{\rm th}}$ stands for the square of the threshold energy.
Now the spectral function is taken as
  \begin{eqnarray}
  {\rm Im}D(s_V)
  =
  {\rm Im} \left\{ \frac{1}{s_V - M_V^2 + iM_V \Gamma_V} \right\}
  \label{eq:im_disp},
  \end{eqnarray}
where the width $\Gamma_V$ is assumed to be a constant physical value.
Substituting eqs.~\eqref{eq:disp} and \eqref{eq:im_disp} into the original loop function eq.~\eqref{eq:G_rhoK}, we have
  \begin{eqnarray}
  \tilde{G}_l(\sqrt{s})
  &=&
  \frac{1}{C_l}
  \int_{(M_l - 2 \Gamma_l)^2}^{(M_l + 2\Gamma_l)^2}
  ds_V G_l (\sqrt{s},\sqrt{s_V},m_l) \nonumber \\
  &\times&
  \left( - \frac{1}{\pi} \right)
  {\rm Im} \left\{ \frac{1}{s_V - M_l^2 + iM_l \Gamma_l} \right\} ,
  \end{eqnarray}
with the normalization for the $l$th component
  \begin{eqnarray}
  C_l
  &=&
  \int_{(M_l - 2 \Gamma_l)^2}^{(M_l + 2\Gamma_l)^2}
  ds_V \times
  \left( - \frac{1}{\pi} \right)
  {\rm Im} \left\{ \frac{1}{s_V - M_l^2 + iM_l \Gamma_l} \right\} , \nonumber \\
  \end{eqnarray}
with $m_l$, $M_l$, $\Gamma_l$, the mass of the pseudoscalar meson, mass of the vector and width of the vector respectively.
Replacing $G_l$ by $\tilde{G}_l$ in eq.~\eqref{eq:rhoK_amp}, we include the width effect of vector mesons.

In order to obtain the $VP$ interaction kernel in terms of the $SU(3)$ chiral symmetry, we start from the following Lagrangian
  \begin{eqnarray}
  {\cal L}
  =
  -\frac{1}{4}\left\langle
  V_{\mu \nu} V^{\mu \nu}
  \right\rangle ,
  \label{eq:L_kin}
  \end{eqnarray}
with the field strength tensor
  \begin{eqnarray}
  V_{\mu \nu}
  &=&
  \nabla_{\mu} V_\nu - \nabla_{\nu} V_\mu ,
  \end{eqnarray}
where $\langle \cdots \rangle$ stands for taking the trace of an inside matrix.
The covariant derivative is determined with the pseudoscalar field
  \begin{eqnarray}
  \nabla_{\mu} V_\nu
  &=&
  \partial_{\mu} V_{\nu} + [ \Gamma_\mu, V_\nu ] ,\\
  \Gamma_\mu
  &=&
  \frac{1}{2} (u^\dagger \partial_\mu u + u \partial_\mu u^\dagger ) , \\
  u^2
  &=&
  U
  =
  \exp \left[i\frac{ \sqrt{2} P}{f} \right] ,
  \end{eqnarray}
where $f$ is the pion decay constant in the $SU(3)$ limit.
In the equations above, the two field matrices of the pseudoscalar $P$ and vector $V$ are given by
  \begin{eqnarray}
  P=
  \left(
  \begin{array}{ccc}
  \frac{1}{\sqrt{2}}\pi^0 + \frac{1}{\sqrt{6}}\eta_8 & \pi^+ & K^{+} \\
  \pi^- & -\frac{1}{\sqrt{2}}\pi^0 + \frac{1}{\sqrt{6}}\eta_8 & K^{0} \\
  K^{-} & \bar{K}^{0} & - \frac{2}{\sqrt{6}}\eta_8 \\
  \end{array}
  \right) ,
  \end{eqnarray}
and
  \begin{eqnarray}
  V_\mu=
  \left(
  \begin{array}{ccc}
  \frac{1}{\sqrt{2}}\rho^0 & \rho^+ & K^{*+} \\
  \rho^- & -\frac{1}{\sqrt{2}}\rho^0+\frac{1}{\sqrt{2}}\omega & K^{*0} \\
  K^{*-} & \bar{K}^{*0} & \phi \\
  \end{array}
  \right)_{\mu} .
  \end{eqnarray}
In the WCCWZ approach \cite{Weinberg:1968de,Coleman:1969sm,Ecker:1989yg}, this Lagrangian stems from a nonlinear realization of chiral symmetry.
Expanding eq.~\eqref{eq:L_kin} up to two pseudoscalar fields, we have the leading order contribution of the four point $VVPP$ interaction Lagrangian
  \begin{eqnarray}
  {\cal L}_{VVPP}
  =
  -\frac{1}{4f^2}
  \left\langle
  \left[ V^\mu, \partial^\nu V_\mu \right]
  \left[ P, \partial_\nu P \right]
  \right\rangle .
  \label{eq:L_VP}
  \end{eqnarray}
From eq.~\eqref{eq:L_VP}, the $VP$ potential projected over $s$-wave, can be obtained as
  \begin{eqnarray}
  V_{ij}(s)
  &=&
  -\frac{\vec{\epsilon}\cdot \vec{\epsilon}'}{8f^2} C_{ij}
  \left[
  3s - (M_i^2 + m_i^2 + M_j^2 + m_j^2)
  \right. \nonumber \\
  &&
  \left.
  - \frac{1}{s}(M_i^2 - m_i^2)(M_j^2 - m_j^2)
  \right] ,
  \label{eq:V_VVPP}
  \end{eqnarray}
where the index $i(j)$ represents the $VP$ channel of the incoming (outgoing) particles.
In order to construct the $VP$ amplitude in the isospin basis, we follow the phase convention shown below for states $| I, I_z \rangle$
  \begin{eqnarray}
  | \pi^+ \rangle = -|1,+1 \rangle &,&~
  | \rho^+ \rangle = -|1,+1 \rangle ,~ \nonumber \\
  | K^- \rangle = -|1/2,-1/2 \rangle &,&~
  | K^{*-} \rangle = -|1/2,-1/2 \rangle .
  \label{eq:phase}
  \end{eqnarray}
The coefficients $C_{ij}$ in eq.~\eqref{eq:V_VVPP} for the $I=1/2$ sector are given and tabulated in table \ref{table:rhoK1/2}.
As it will be discussed later, we also need the $VP$ potential in the $I=3/2$ sector which can be obtained in the same manner from eq.~\eqref{eq:L_VP}.
In the $I=3/2$ sector, the relevant channels are $\rho K$ and $K^* \pi$ and the coefficients are given in table~\ref{table:rhoK3/2}.
In order to have an appropriate unitarized amplitude, we use the following parameter set chosen to reproduce $K_1 (1270)$ in ref.~\cite{Geng:2006yb}
  \begin{eqnarray}
  \mu=900~{\rm MeV},~ a(\mu)=-1.85,~f=115~{\rm MeV}.
  \end{eqnarray}

   \begin{table}
   \caption{ Coefficients $C_{ij}$ in eq.~\eqref{eq:V_VVPP} in the $I=1/2$ sector.}
   \begin{center}
   \begin{tabular}{c|ccccc}
   \hline
   \hline
                & $\phi K$ & $\omega K$ & $\rho K$ & $K^* \eta$ & $K^* \pi$ \\
   \hline
   $\phi K$     & 0 & 0 & 0 & $-\sqrt{\frac{3}{2}}$ & $-\sqrt{\frac{3}{2}}$ \\
   $\omega K$   & 0 & 0 & 0 & $\frac{\sqrt{3}}{2}$ & $\frac{\sqrt{3}}{2}$ \\
   $\rho K$     & 0 & 0 & $-2$ & $-\frac{3}{2}$ & $ \frac{1}{2}$ \\
   $K^* \eta$   & $-\sqrt{\frac{3}{2}}$ & $\frac{\sqrt{3}}{2}$ & $-\frac{3}{2}$ & 0 & 0\\
   $K^* \pi$    & $-\sqrt{\frac{3}{2}}$ & $\frac{\sqrt{3}}{2}$ & $\frac{1}{2}$ & 0 & $-2$ \\
   \hline
   \hline
   \end{tabular}
   \label{table:rhoK1/2}
   \end{center}
   \end{table}

   \begin{table}
   \caption{ Coefficients $C_{ij}$ in eq.~\eqref{eq:V_VVPP} in the $I=3/2$ sector.}
   \begin{center}
   \begin{tabular}{c|cc}
   \hline
   \hline
              & $\rho K$ & $K^* \pi $\\
   \hline
   $\rho K$   & 1 & 1 \\
   $K^* \pi$  & 1 & 1 \\
   \hline
   \hline
   \end{tabular}
   \label{table:rhoK3/2}
   \end{center}
   \end{table}

\section{\label{sec:rhoKKbar} The $\rho K\bar{K}$ three-body scattering}

Once the unitarized $\rho K$ amplitude is obtained, let us go to the $\rho K\bar{K}$ three-body system.
As mentioned above, we study this system by solving the Faddeev equation within the fixed center approximation.
We follow refs. \cite{Roca:2010tf,YamagataSekihara:2010qk,Xiao:2012dw,Liang:2013yta}, and here we provide only a brief description of the formalism.

\begin{center}
  \begin{figure}
  \resizebox{0.5\textwidth}{!}{%
  \includegraphics{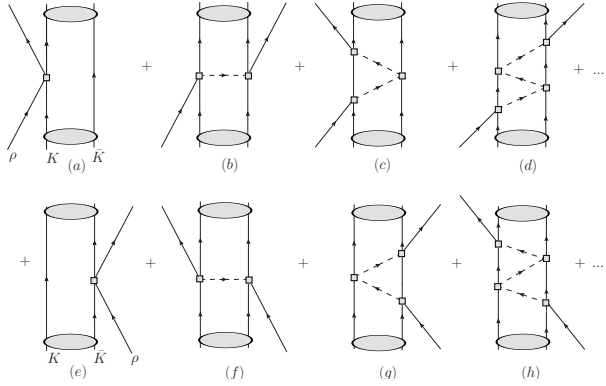}
  }
  \caption{ \label{fig:fca}
  Diagrammatic representation of the fixed center approximation for the $\rho K \bar{K}$ system.}
  \end{figure}
  \end{center}

Under the fixed center approximation, it is assumed that two particles 1 and 2 cluster together and the structure of the cluster is kept against the collision of the extra particle 3.
This idea leads the three-body scattering amplitude $T$ which can read a summation of the two following partition functions $T_1$ and $T_2$
  \begin{eqnarray}
  \label{eq:fca}
  T_1
  &=&
  t_1 + t_1 G_0 T_2, \nonumber \\
  T_2
  &=&
  t_2 + t_2 G_0 T_1, \\
  T
  &=&
  T_1 + T_2 \nonumber ,
  \end{eqnarray}
where the subscripts $i=1,2$ of $T_i$ and $t_i$ represent the component particle $i$ in the cluster and the diagrammatic sketches are depicted in fig.~\ref{fig:fca}.
The interaction kernel $t_i$ between the particle $i$ and particle 3 stands for the leading order contribution in the partition function $T_i$ (see the (a) or (e) in fig.~\ref{fig:fca}) .
From the double scattering, the propagation of the particle 3 appears and is described as a function $G_0$.
In the present work, the particles 1, 2 and 3 correspond to $K$, $\bar{K}$ and $\rho$ respectively, and thus $t=t_1=t_2$.
Therefore, from now on we will not specify the particle 1 or 2.

In order to obtain the scattering amplitude, following refs.~\cite{YamagataSekihara:2010qk,Xiao:2012dw,Liang:2013yta}, we determine the field normalization.
From eq.~\eqref{eq:fca}, the $S$-matrix for the single scattering which corresponds to $(a)$ or $(e)$ in fig.~\ref{fig:fca} can be derived as
  \begin{eqnarray}
  S^{(1)}
  &=&
  -i(2\pi)^4 \delta^4 (k_{\rho} + k_{f_0} - k_{\rho}^{'} - k_{f_0}^{'}) \nonumber \\
  &\times&
  \frac{1}{{\cal V}^2}
  \frac{1}{\sqrt{2\omega_\rho}} \frac{1}{\sqrt{2\omega_\rho^{'}}}
  \frac{1}{\sqrt{2\omega_K}} \frac{1}{\sqrt{2\omega_K^{'}}} t,
  \label{eq:S_mat1}
  \end{eqnarray}
where the momentum $k(k')$, the on-shell energy $\omega (\omega^{'})$ are assigned to the initial (final) particles respectively and ${\cal V}$ is the volume of the box to normalize the external fields to unity.
There the assumption of the cluster $f_0 (980)$ is used.
In the same manner, we have the $S$-matrix for the double scattering, $(b)$ or $(f)$ in fig.~\ref{fig:fca}
  \begin{eqnarray}
  S^{(2)}
  &=&
  -i(2\pi)^4 \delta^4 (k_{\rho} + k_{f_0} - k_{\rho}^{'} - k_{f_0}^{'}) \nonumber \\
  &\times&
  \frac{1}{ {\cal V}^2 }
  \frac{1}{ \sqrt{2\omega_\rho} } \frac{1}{ \sqrt{2\omega_\rho^{'}} }
  \frac{1}{ \sqrt{2\omega_K} } \frac{1}{ \sqrt{2\omega_K^{'}} }
  \frac{1}{ \sqrt{2\omega_K} } \frac{1}{ \sqrt{2\omega_K^{'}} } \nonumber \\
  &\times&
  \int \frac{d^3 q}{(2\pi)^3} F_{f_0}(q) \frac{1}{ q^{02} -\vec{q}^2 - m_{\rho}^2 + i\epsilon } tt,
  \label{eq:S_mat2}
  \end{eqnarray}
where the wave function of the $K \bar{K}$ bound state is modeled by the form factor $F_{f_0}$ and it will be discussed below.
The integration which appears in eq.~\eqref{eq:S_mat2}, in general, should be carried out with a four dimensional momentum.
However, by the help of the on-shell factorization, $t$ can be factorized out of the integration and one is left with a three dimensional integration.
Thus the $G_0$ function of eq.~\eqref{eq:fca} reads as a function of the energy $\sqrt{s}$
  \begin{eqnarray}
  G_0(\sqrt{s})
  =
  \frac{1}{2M_{f_0}}\int
  \frac{d^3q}{(2\pi)^3}
  F_{f_0}(q) \frac{1}{q^{02}(s) - \vec{q}^2 - m_{\rho}^2 + i\epsilon} , \nonumber \\
  \label{eq:G0}
  \end{eqnarray}
where $M_{f_0}$ is the mass of the $f_0 (980)$ resonance.
The energy of the propagator $q^0$ is determined at the three-body rest frame
  \begin{eqnarray}
  q^0(\sqrt{s})
  =
  \frac{ s+m_\rho^2-M_{f_0}^2 }{2\sqrt{s}}.
  \end{eqnarray}
Besides it should be noted here that the unitarized amplitude $t$ is a function of $\sqrt{s'}$ instead of $\sqrt{s}$ as discussed in ref.~\cite{YamagataSekihara:2010qk}
  \begin{eqnarray}
  s'= \frac{1}{2} (s + M_{\rho}^2 + 2 m_{K}^2 - M_{f_{0}}^2) .
  \end{eqnarray}
The full three-body scattering is given by
  \begin{eqnarray}
  S
  &=&
  -i(2\pi)^4 \delta^4 (k_{\rho} + k_{f_0} - k_{\rho}^{'} - k_{f_0}^{'}) \nonumber \\
  &=&
  \frac{1}{{\cal V}^2}
  \frac{1}{ \sqrt{ 2\omega_\rho } } \frac{1}{ \sqrt{ 2\omega_\rho^{'} } }
  \frac{1}{ \sqrt{ 2\omega_{f_0}} } \frac{1}{ \sqrt{ 2\omega_{f_0}^{'}} } T .
  \label{eq:S_mat}
  \end{eqnarray}
Therefore combining eqs.~\eqref{eq:S_mat1}, \eqref{eq:S_mat2}, \eqref{eq:G0} and \eqref{eq:S_mat}, the three-body scattering amplitude is described as a series expansion.
Besides, using the low energy reduction, $\sqrt{2\omega} \sim \sqrt{2m}$, we have a simple expression
  \begin{eqnarray}
  T(s)
  &=&
  2 \left[ \tilde{t}(s') + \tilde{t}(s') G_0(s) \tilde{t}(s') + \cdots \right] \nonumber \\
  &=&
  2 \frac{ \tilde{t}(s')}{1 - \tilde{t}(s') G_0(s)},
  \label{eq:amp_rhoKK}
  \end{eqnarray}
where $t$ is replaced by $\tilde{t}$ with a factor coming from the normalization of the fields
  \begin{eqnarray}
  \tilde{t}
  =
  \frac{2m_{f_0}}{2m_K} t .
  \end{eqnarray}

\begin{center}
  \begin{figure}
  \resizebox{0.5\textwidth}{!}{%
  \includegraphics{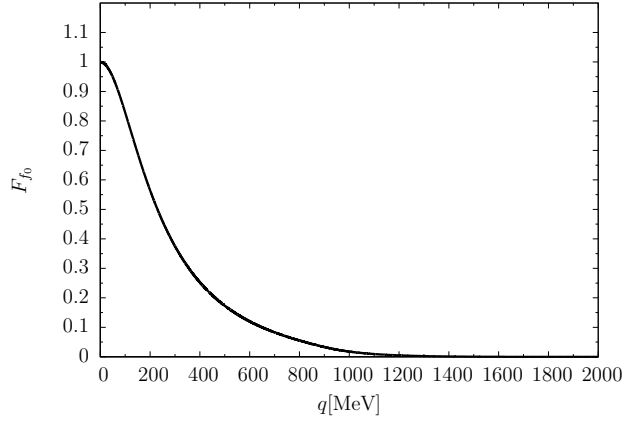}
  }
  \caption{ \label{fig:ff}
  The form factor for the $K\bar{K}$ cluster given by eq.~\eqref{eq:ff}.}
  \end{figure}
  \end{center}

\begin{center}
  \begin{figure}
  \resizebox{0.5\textwidth}{!}{%
  \includegraphics{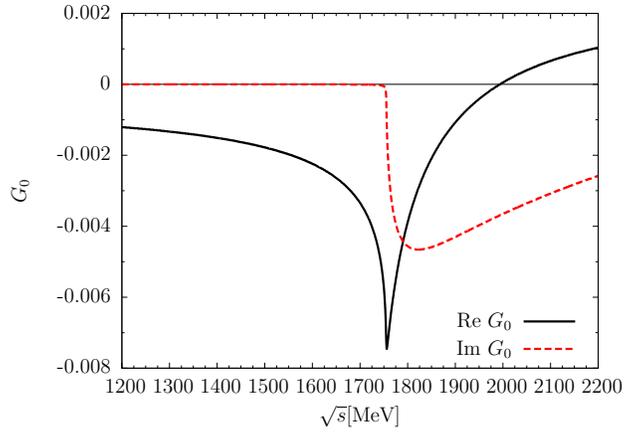}
  }
  \caption{ \label{fig:G0}
  The $G_0$ function given by eq.~\eqref{eq:G0}.}
  \end{figure}
  \end{center}

Upon implementing the Faddeev equations under the fixed center approximation, we have to take into account the information of the cluster appropriately.
Therefore the wave function of bound states or resonances of the component particles in the cluster should be considered.
In order to relate the wave function with dynamically generated states, for arbitrary angular momentum, the formalism has been developed through refs.~\cite{Gamermann:2009uq,YamagataSekihara:2010pj,Aceti:2012dd}.
We apply it to the $K \bar{K}$ $s$-wave bound state and provide only the expression to the form factor
  \begin{eqnarray}
  F_{f_0} (q)
  &=&
  \frac{1}{{\cal N}}
  \int_{ \stackrel{p < k_{{\rm max} }}{|{\vec{p} - \vec{q}}| < k_{{\rm max}}}}
  d^3p
  \left( \frac{1}{2 \omega_K(\vec{p})}\right)^2
  \frac{1}{M_{f_0}- 2 \omega_K (\vec{p})} \nonumber \\
  &\times&
  \left( \frac{1}{2 \omega_K(\vec{p} -\vec{q})}\right)^2
  \frac{1}{M_{f_0}- 2 \omega_K ( \vec{p}-\vec{q}) },
  \label{eq:ff}
  \end{eqnarray}
where the normalization ${\cal N}$ is given by
  \begin{eqnarray}
  {{\cal N}}
  =
  \int_{p < k_{{\rm max}}} d^3p \left[ \left( \frac{1}{2\omega_K (\vec{p})}\right)^2 \frac{1}{M_{f_0}- 2 \omega_K ( \vec{p} )} \right]^2 . \nonumber \\
  \label{eq:norm_ff}
  \end{eqnarray}
From ref.~\cite{Oller:1997ti}, we take $k_{\rm max}=\sqrt{\Lambda^2 - m_K^2}$ and $\Lambda = 1030$ MeV for getting the $f_0 (980)$ from the $K \bar{K}$ cluster.
Note that in the region $q > 2k_{\rm max}$, $F_{f_0}$ vanishes identically.
Therefore the integration in eq.~\eqref{eq:G0} has a limit $2 k_{\rm max}$.
In figs.~\ref{fig:ff} and \ref{fig:G0}, the form factor for the $f_0(980)$ resonance and the $G_0$ function are shown, respectively.
In the form factor the strong suppression of high energy momentum is seen and the threshold effect appears in the $G_0$ function.

At the end of this section, we refer to the conversion of the basis of the unitarized amplitude $t$.
Since the $\rho [K\bar{K}]_{I=0} $ system is studied in the sector $I(J^P)=1(1^-)$, now we focus on the scattering in the $I_z=+1$ channel, $\rho^+ [K \bar{K}]_{I=0}$.
With the phase convention given in eq.~\eqref{eq:phase}, the $K \bar{K}$ field is given by
  \begin{eqnarray}
  | K \bar{K} \rangle_{I=0}
  =
  - \frac{1}{\sqrt{2}}
  \left(
  \left| K^+ K^- \right\rangle + \left| K^0 \bar{K}^0 \right\rangle
  \right) .
  \end{eqnarray}
So far we consider the single scattering contribution which corresponds to $(a)$ in fig.~\ref{fig:fca}.
Following the same phase convention, we have
  \begin{eqnarray}
  t
  &=&
  \left\langle \rho^+ \right| \otimes \left\langle [K \bar{K}]_{I=0} \right|
  \hat{t}_{\rho K}
  \left| [K \bar{K}]_{I=0} \right\rangle \otimes \left| \rho^+ \right\rangle \nonumber \\
  &=&
  \frac{1}{2}
  \left(
  \left\langle \rho^+ K^+ K^- \right| + \left\langle \rho^+ K^0 \bar{K}^0 \right|
  \right) \hat{t}_{\rho K} \nonumber \\
  &\times&
  \left(
  \left| \rho^+ K^+ K^- \right\rangle + \left| \rho^+ K^0 \bar{K}^0 \right\rangle
  \right) \nonumber \\
  &=&
  \frac{1}{2}
  \left(
    \left\langle \rho^+ K^+ \right| \hat{t}_{\rho K} \left| \rho^+ K^+ \right\rangle
  + \left\langle \rho^+ K^0 \right| \hat{t}_{\rho K} \left| \rho^+ K^0 \right\rangle
  \right), \nonumber \\
  \end{eqnarray}
where $\bar{K}$ acts as a spectator.
Therefore the interaction kernel is given as a mixture of different isospin sectors
  \begin{eqnarray}
  t
  =
  \frac{1}{3} \left( 2 t_{\rho K}^{I=3/2} + t_{\rho K}^{I=1/2} \right) ,
  \end{eqnarray}
where $t_{\rho K}$ is the $\rho K$ unitarized scattering amplitude given by eq.~\eqref{eq:rhoK_amp}.

\section{\label{sec:results} Results}

\begin{center}
  \begin{figure}
  \resizebox{0.5\textwidth}{!}{%
  \includegraphics{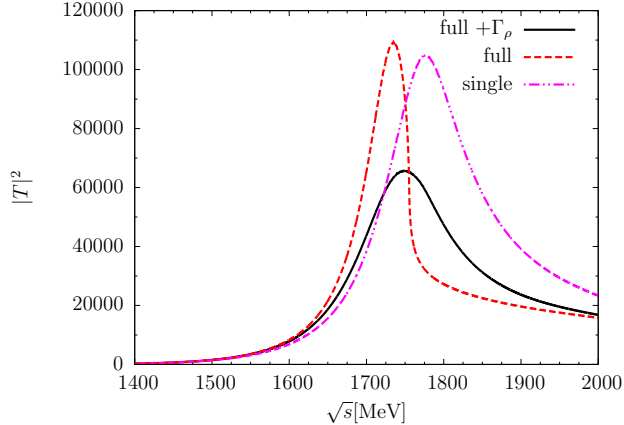}
  }
  \caption{The $\rho K\bar{K}$ amplitude where ``single", ``full", ``full + $\Gamma_{\rho}$" denote the amplitude of the single scattering, full scattering, full scattering with the $\rho$ width effect, respectively.}
  \label{fig:amp}
  \end{figure}
  \end{center}

Now we have the $\rho K \bar{K}$ three-body amplitude within the fixed center approximation.
To see the effect of the multiple scattering, the full amplitude is compared with the single scattering amplitude $T=2\tilde{t}$.
Furthermore, as discussed in the previous work~\cite{Geng:2006yb}, the inclusion of the vector meson width is found to be much important.
Therefore, we also consider the width of the $\rho$, $\Gamma_\rho \sim$ 150 MeV, in the $G_0$ function.
By using the following replacement for the $\rho$ propagator in eq.~\eqref{eq:G0}
  \begin{eqnarray}
  \frac{1}{q^{02}-\vec{q}^2 - m_{\rho}^2 + i\epsilon}
  \to
  \frac{1}{q^{02}-\vec{q}^2 - m_{\rho}^2 + i m_{\rho} \Gamma_{\rho}},
  \end{eqnarray}
we include the $\rho$ width into the $G_0$ function.
In fig.~\ref{fig:amp}, the $\rho^+ [K\bar{K}]_{I=0}$ amplitude is shown and there one can see that a peak appears in each case.
In the single scattering amplitude, the peak exists above the threshold of $\rho$ and $f_0 (980)$.
Through the multiple scattering, the peak position shifts lower while the width is getting smaller because the $\rho f_0(980)$ channel becomes closed.
With the $\rho$ width effect, that peak is getting much wider while the shift of the peak position is not so large.
The masses (or a peak position in the amplitude) and full widths at half maximum of the dynamically generated state are listed with the experimental data in table.~\ref{table:amp}.
Compared with the experimental data, it is found to be important to take into account the $\rho$ width.

   \begin{table}
   \caption{The masses and widths of dynamically generated states where the amplitudes are given by the single scattering, the full scattering and the full scattering with the $\rho$ width effect respectively.}
   \begin{center}
   \begin{tabular}{c|cccc}
   \hline
   \hline
              & single & full & full + $\Gamma_{\rho}$ & PDG \cite{Beringer:1900zz} \\
   \hline
   Mass (MeV) & 1777.9 & 1734.8 & 1748.0 & 1720 $\pm$ 20 \\
   Width (MeV) & 144.4  & 63.7   & 160.8  & 250 $\pm$ 100 \\
   \hline
   \hline
   \end{tabular}
   \label{table:amp}
   \end{center}
   \end{table}

\begin{center}
  \begin{figure}
  \resizebox{0.5\textwidth}{!}{
  \includegraphics{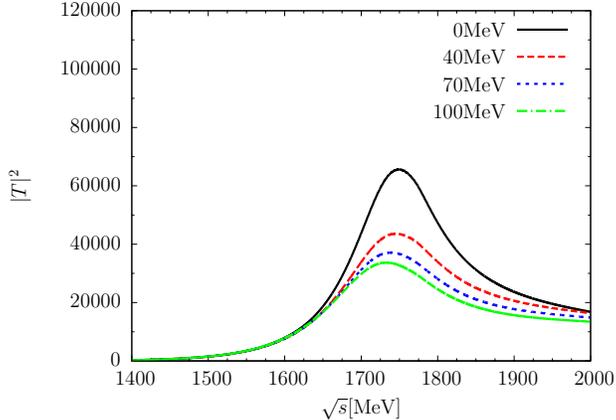}
  }
  \caption{The $\rho K\bar{K}$ amplitude with the $\rho$ and $f_0(980)$ width effect, taking $\Gamma_{f_0}$ as 0, 40, 70 and 100 MeV, respectively.}
  \label{fig:amp_width}
  \end{figure}
  \end{center}

As shown above, the present study seems to work for generating the resonance which might correspond to $\rho(1700)$.
In addition, we consider the width of the $f_0 (980)$ too.
The $K\bar{K}$ component in the $f_0 (980)$ resonance is found to be dominant while the decay width into the $\pi \pi$ channel is not so small.
Therefore we consider the inclusion of the width into our framework.
In order to keep the fixed center approximation, here we give a naive prescription that the eigenvalue of the $K \bar{K}$ system is now a complex value.
Namely the mass of the cluster $M_{f_0}$ in eqs. \eqref{eq:ff} and \eqref{eq:norm_ff} is replaced by $ M_{f_0} - i\Gamma_{f_0} /2$.
Then the form factor becomes a complex function which might represent the effect of the wave function of the unstable cluster.
The amplitude with the $f_0(980)$ and $\rho$ width effect is shown in fig.~\ref{fig:amp_width} and the masses and widths of the dynamically generated state are listed in table~\ref{table:amp_width}.
Taking into account the ambiguity of the $f_0 (980)$ width, as a value of $\Gamma_{f_0}$, a maximum and minimum value of the experimental data and their average are taken.
It is shown that the inclusion of the $f_0(980)$ width induces a suppression of the magnitude of the peak and the peak becomes broader as the width of the $f_0 (980)$ increases.
Furthermore it is also a remarkable feature that the peak position is not so affected by this prescription.
By taking $\Gamma_{f_0}=70$ MeV which is between 40 MeV and 100 MeV as quoted in the PDG\cite{Beringer:1900zz}, we find the mass at 1739 MeV and the width of 227 MeV which agrees very well with the experimental properties of the $\rho (1700)$ (see table~\ref{table:amp}) . 

  \begin{table}
  \caption{The masses and widths of dynamically generated states with the $\rho$ and $f_0(980)$ width effects. (in MeV)}
  \begin{center}
  \begin{tabular}{c|cccc}
  \hline
  \hline
              & $\Gamma_{f_0}=0$ & $\Gamma_{f_0}=40$ & $\Gamma_{f_0}=70$ & $\Gamma_{f_0}=100$ \\
  \hline
  Mass   & 1748.0 & 1743.6 & 1739.2 & 1734.8 \\
  Width  &  160.8 &  216.4 &  227.2 &  224.6 \\
  \hline
  \hline
  \end{tabular}
  \end{center}
  \label{table:amp_width}
  \end{table}

\section{conclusion}
Through the present paper, we construct the $\rho K\bar{K}$ three-body amplitude by means of the fixed center approximation.
In our framework, a pair of $K \bar{K}$ is considered to form a scalar meson cluster $f_0 (980)$, based on ref.~\cite{Oller:1997ti}.
We use the $\rho K$ unitarized amplitude provided by refs.~\cite{Roca:2005nm,Geng:2006yb} in a manner giving a respect to chiral symmetry.
In the three-body amplitude, we have a peak at the energy around 1748 MeV rather independent of the width of the $f_0 (980)$.
Besides, it is seen that the inclusion of the $\rho$ and $f_0 (980)$ width makes the peak wider and gives a good agreement with the experimental data of the $\rho (1700)$, both for the position and the width.
Since the $\rho$ decays into $\pi \pi$ mostly, the above results might be related to the dominant decay mode of the $\rho (1700)$, $\rho \pi \pi$ and 4$\pi$.
Our approach to the $\rho K \bar{K}$ system provides the description of the $\rho (1700)$ as a dynamically generated state and then we conclude that the building block of the $\rho (1700)$ resonance are the $\rho$ and $ f_0(980)$.

\section*{Acknowledgments}  
We would like to thank L.~S.~Geng, E.~Oset, L.~Roca and J.~A.~Oller for providing us the pseudoscalar-vector amplitude.
This work is partly supported by the Spanish Ministerio de Economia y Com- petitividad and European FEDER funds under the Contract No. FIS2011-28853-C02-01 and the Generalitat Valenciana in the program Prometeo, 2009/090. We acknowledge the support of the European Community-Research Infrastructure Integrating Activity Study of Strongly Interacting Matter(acronym Hadron Physics 3, Grant No. 283286) under the Sev- enth Framework Programme of the European Union.
This work is also partly supported by the National Natural Science Foundation of China under Grant No. 11165005 and by scientific research fund (201203YB017) of the Education Department of Guangxi.

\end{document}